\documentclass[aps,prd]{revtex4}
\newcommand{\slashs}[1]{\not{\!#1}}
\usepackage{epsfig}
\begin{document}

\title{$B$ meson wave function from the $B\to\gamma
l\nu$ decay}
\author{Yeo-Yie Charng$^1$}
\email{charng@phys.sinica.edu.tw}
\author{Hsiang-nan Li$^{1,2}$}
\email{hnli@phys.sinica.edu.tw}
\affiliation{$^1$Institute of Physics, Academia Sinica, Taipei,\\
Taiwan 115, Republic of China}
\affiliation{$^2$Department of Physics, National Cheng-Kung University,\\
Tainan, Taiwan 701, Republic of China}

\begin{abstract}

We show that the leading-power $B$ meson wave function can be
extracted reliably from the photon energy spectrum of the
$B\to\gamma l\nu$ decay up to $O(1/m_b^2)$ and $O(\alpha_s^2)$
uncertainty, $m_b$ being the $b$ quark mass and $\alpha_s$ the
strong coupling constant. The $O(1/m_b)$ corrections from
heavy-quark expansion can be absorbed into a redefined
leading-power $B$ meson wave function. The two-parton $O(1/m_b)$
corrections cancel exactly, and the three-parton $B$ meson wave
functions turn out to contribute at $O(1/m_b^2)$. The constructive
long-distance contribution through the $B\to V\to\gamma$
transition, $V$ being a vector meson, almost cancels the
destructive $O(\alpha_s)$ radiative correction.
%The effect from bremsstrahlung photon
%emission vanishes like the lepton mass due to helicity
%suppression.
Using models of the leading-power $B$ meson wave function
available in the literature, we obtain the photon energy spectrum
in the perturbative QCD framework, which is then compared with
those from other approaches.

\end{abstract}
\maketitle

\section{INTRODUCTION}

The two-parton leading-power (LP) $B$ meson wave function
(distribution amplitude) $\phi_+$ plays an essential role in a
perturbative analysis of exclusive $B$ meson decays based on $k_T$
factorization theorem \cite{BS,LS,NL2} (collinear factorization
theorem \cite{BL,ER,CZS,CZ,L1,AVR}). Its behavior certainly
matters, and has been investigated in various approaches recently.
Models of the distribution amplitude $\phi_+(x)$ with an
exponential tail in the large $x$ region have been proposed
\cite{GN}, where $x$ is the longitudinal momentum fraction carried
by the light spectator quark. Neglecting three-parton distribution
amplitudes in a study by means of equations of motion
\cite{PB1,Braun:1990iv}, $\phi_+(x)$ was found to be proportional
to a step function with a sharp drop at large $x$ \cite{KKQT}. The
wave function $\phi_+(x,k_T)$, where $k_T$ is the transverse
momentum carried by the light spectator quark, was also derived in
the same framework \cite{KKQT}. All these models depend on at
least one shape parameter, whose determination requires
experimental inputs from exclusive $B$ meson decays.

%The $B\to\gamma l\nu$ decays, $l=e$, $\mu$,  Different approaches,
%such as light-cone sum rules (LCSR), perturbative QCD (PQCD),
%quark model, light-front QCD (LFQCD), the Bethe-Salpeter
%equation,..., have been applied , in exclusive $B$ meson decays.
%Even such a simple mode involves abundant dynamics.

In this paper we shall show that the radiative decay $B\to\gamma
l\nu$ provides the cleanest information of the LP $B$ meson wave
function $\phi_+$. This mode has been widely studied in
\cite{NL2,L1,PRD51-111,PLB358-329,PLB361-137,PLB372-331,
MPLA11-1061,PRD57-5697,PRD61-114510,
PRD64-097503,NPB649-349,BHLN,NPB650-356,JHEP042003029,KAG,NL,L5}
due to different motivations: for extracting the $B$ meson decay
constant $f_B$ and the Cabibbo-Kobayashi-Maskawa (CKM) matrix
element $|V_{ub}|$, for demonstrating the next-to-leading-order
(NLO) calculation and the proof of QCD factorization theorem, for
deriving resummation of large logarithmic corrections, for
studying long-distance effect and annihilation mechanism,.... The
subject on the extraction of the $B$ meson wave function from the
$B\to\gamma l\nu$ data has not yet been discussed. It will be
shown that two-parton next-to-leading-power (NLP) [$O(1/m_b)$]
corrections cancel exactly, $m_b$ being the $b$ quark mass. The
contributions from higher Fock states, the three-parton $B$ meson
wave functions, turn out to be of $O(1/m_b^2)$. The constructive
long-distance contribution through the $B\to V\to\gamma$
transition, $V$ being a vector meson, almost cancels the
destructive $O(\alpha_s)$ radiative correction, $\alpha_s$ being
the strong coupling constant. The effect from bremsstrahlung
photon emissions vanishes like the lepton mass because of helicity
suppression. Therefore, the extraction of $\phi_+$ from the
measured photon energy spectrum of the $B\to\gamma l\nu$ decay
suffers only $O(1/m_b^2)$ and $O(\alpha_s^2)$ uncertainty.

We identify and discuss the higher-power corrections to the
$B\to\gamma l\nu$ decay in Sec.~II, and calculate the long- and
short-distance effects in Sec.~III. Section IV is the conclusion.
The hard kernel associated with the three-parton distribution
amplitudes is derived in the Appendix, whose explicit expression
is necessary for demonstrating the smallness of the
higher-Fock-state contribution. Our conclusion differs from that
drawn in \cite{KMO0504}, in which the semileptonic decay $B\to\pi
l\nu$ was regarded as a more ideal process for extracting the $B$
meson wave function. The argument is that the radiative decay
$B\to\gamma l\nu$, receiving a large long-distance uncertainty,
does not serve the purpose. As stated above, this long-distance
effect is in fact cancelled by the $O(\alpha_s)$ short-distance
one almost exactly.

\section{HIGHER-POWER CORRECTIONS}

In this section we identify and discuss higher-power corrections
to the $B\to\gamma l\nu$ decay. The $B$ meson momentum $P_1$ and
the photon momentum $P_2$ are parameterized, in the light-cone
coordinates, as
\begin{eqnarray}
P_1=\frac{m_B}{\sqrt{2}}(1,1,{\bf 0}_T)\;,\;\;\;
P_2=\frac{m_B}{\sqrt{2}}(0,\eta,{\bf 0}_T)\;, \label{bmpp}
\end{eqnarray}
respectively, where $\eta\equiv 2E_\gamma/m_B$, $m_B$ being the
$B$ meson mass, denotes the photon energy fraction. The decay
amplitude is decomposed into
\begin{eqnarray}
\frac{1}{e}\left\langle \gamma \left( P_2,\epsilon_T\right) \left|
\bar{u}\gamma_\mu (1-\gamma _5)b\right| \bar B\left( P_1\right)
\right\rangle =\epsilon_{\mu \nu \alpha \beta }\epsilon_T
^{*\nu}v^{\alpha} P_2^\beta F_V(q^2)+i\left[\epsilon ^*_{T\mu}
\left( v\cdot P_2\right)- \left( \epsilon_T ^{*}\cdot v\right)
P_{2\mu} \right] F_A(q^2)\;, \label{ME1b}
\end{eqnarray}
where $e$ is the electron charge, $\epsilon_T$ the polarization
vector of the photon, $v=P_1/m_B$ the $B$ meson velocity, and
$q^2\equiv (P_1-P_2)^2=(1-\eta)m_B^2$ the lepton-pair invariant
mass. The decay spectrum is then given, in terms of the form
factors $F_{V,A}$, by
\begin{eqnarray}
\frac{d\Gamma }{d\eta}&=&\frac{\alpha G_F^2\left| {V}_{ub}\right|
^2 }{96\pi ^2}m_B^5
(1-\eta)\eta^3\left[F_V^2(q^2)+F_A^2(q^2)\right]\;,\label{d11}
\end{eqnarray}
with $\alpha\equiv e^2/(4\pi)$ and the Fermi constant $G_F$.

\begin{figure}[t!]
\begin{center}
\epsfig{file=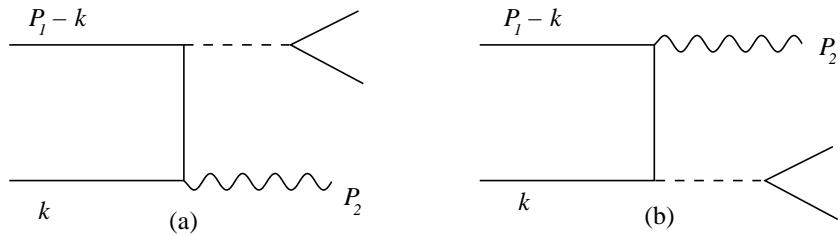,height=1.2in}
\end{center}
\caption{Lowest-order diagrams for the $B\to\gamma l\nu$ decay.
\label{fig1}}
\end{figure}

The collinear factorization theorem for the form factors $F_{V,A}$
in the large $\eta$ region has been proved in
\cite{NPB650-356,NL}, which are expressed as the convolution of
hard kernels with the $B$ meson distribution amplitudes in the
momentum fractions $x$ of the light spectator quark. A hard
kernel, being infrared-finite, is calculable in perturbation
theory. The $B$ meson distribution amplitudes, collecting the soft
dynamics in exclusive $B$ meson decays, are not calculable but
universal. In the framework of factorization theorem, there are
four sources of higher-power corrections to the $B\to\gamma l\nu$
decay:
\begin{enumerate}
\item The heavy-quark expansion of the heavy-light current in
Eq.~(\ref{ME1b}),
\begin{eqnarray}
\bar u\gamma_\mu (1-\gamma _5) b\to \bar u\gamma_\mu (1-\gamma _5)
h+\frac{1}{2m_b}\bar u\gamma_\mu (1-\gamma _5)i\slashs D
h+O(1/m_b^2)\;,\label{cur}
\end{eqnarray}
where the operator $D$ represents the covariant derivative, and
the rescaled $b$ quark field $h$ is related to the full field $b$
by
\begin{eqnarray}
h(z)=\frac{1+\slashs v}{2}e^{i m_bv\cdot z}b(z)\;.
\end{eqnarray}
The factorization of the transition matrix element associated with
the first (second) term in the above expansion leads to the LP
(NLP) $B$ meson distribution amplitudes.

\item The higher-power interactions in the Lagrangian of the
heavy-quark effective theory (HQET). The insertion of the HQET
interactions,
\begin{eqnarray}
O_1=\frac{1}{m_b}\bar h(iD)^2h\;,\;\;\;O_2=\frac{g}{2m_b}\bar
h\sigma^{\mu\nu}G_{\mu\nu}h\;,\label{o12}
\end{eqnarray}
into the transition matrix element associated with the first term
in Eq.~(\ref{cur}) yields $O(1/m_b)$ corrections. We mention that
there exists an alternative heavy-quark effective theory, in which
the higher-power corrections are formulated in a different way
\cite{WWZ03}.

\item The higher Fock states of the $B$ meson. The nonlocal matrix
element,
\begin{eqnarray}
\langle 0 | \bar{u}(z) g G_{\alpha\beta} (uz) h(0) | \bar{B}(P_1)
\rangle\;,\label{m3}
\end{eqnarray}
defines the three-parton distribution amplitudes, where
$G_{\alpha\beta}(uz)$ is the gluon field strength evaluated at the
coordinate $uz$, $0\le u\le 1$. The additional valence gluon,
attaching internal off-shell quark lines, introduces one more hard
propagator, i.e., one more power of $1/m_b$.

\item The subleading parton-level diagrams (hard kernels). The
two-parton lowest-order hard kernels are displayed in Fig.~1,
where the upper quark line represents a $b$ quark. It is easy to
observe that Fig.~\ref{fig1}(a) [(b)] represents the LP (NLP) hard
kernel, since the internal quark line is off-shell by
$m_b\bar\Lambda$ ($m_b^2$) with $\bar\Lambda$ being a hadronic
scale, such as the mass difference $m_B-m_b$.

%\item Power corrections induced by infrared renormalons, which may
%start with $1/m_b^2$. It has been shown \cite{EG} that the leading
%infrared renormalons from the Sudakov resummation and from the
%pole mass of the $b$ quark cancel in inclusive $B$ meson decays.
%The renormalon ambiguity then starts with $1/m_b^2$. Though there
%is a lack of the similar investigation in exclusive $B$ meosn
%decays, it is likely that the conclusion in \cite{EG} applies to
%the $B\to\gamma l\nu$ mode.
\end{enumerate}
%We shall discuss the above sources in the $B\to \gamma l\nu$ decay
%spectrum, and show that they turn out to appear at $1/m_b^2$.

\subsection{Heavy-quark Expansion}

The factorization of soft dynamics from the transition matrix
element associated with the first term on the right-hand side of
Eq.~(\ref{cur}),
\begin{eqnarray}
\left\langle \gamma \left( P_2,\epsilon_T\right) \left|
\bar{u}\gamma_\mu (1-\gamma _5)h\right| \bar B\left( P_1\right)
\right\rangle\;,\label{lm}
\end{eqnarray}
leads to the nonlocal matrix element \cite{GN},
\begin{eqnarray}
\int \frac{dz^-d^2z_T}{(2\pi)^3}e^{i(k^+z^--{\bf k}_T\cdot {\bf
z}_T)} \langle 0|{\bar u}_{\rho}(z)h_{\delta}(0)|\bar
B(P_1)\rangle =i\frac{f_B}{\sqrt{2}}\left\{(\slashs
P_1+m_B)\gamma_5 \left[\slashs n_+\Phi_{+}(k) +\slashs
n_-\Phi_{-}(k)\right] \right\}_{\delta\rho}\;,
%\nonumber\\
%&=&\frac{-i}{\sqrt{2N_c}}\left\{(\slashs P_1+m_B)\gamma_5
%\left[\phi_B(k)+\frac{\slashs n_-}{\sqrt{2}} {\bar
%\phi}_B(k)\right]\right\}_{\delta\alpha}\;,
\label{bwp2}
\end{eqnarray}
which defines the two-parton LP $B$ meson wave functions
$\Phi_\pm$, with the null vectors $n_+=(1,0,0_T)$ and
$n_-=(0,1,0_T)$, and the light quark momentum $k$.
%, and the wave functions,
%\begin{eqnarray}
%\phi_B=\phi_{+}\;,\;\;\; {\bar\phi}_B=\phi_{+}-\phi_{-}\;.
%\end{eqnarray}
Because the photon momentum $P_2$ has been chosen in the minus
direction, the hard kernels for the form factors $F_{V,A}$ are
independent of the component $k^-$, which becomes irrelevant. We
construct the $B$ meson distribution amplitudes $\phi_\pm(x)$,
$x\equiv k^+/P_1^+$, from the $B$ meson wave functions
$\phi_\pm(x,k_T)\equiv P_1^+\Phi_\pm(xP_1^+,k_T)$ by integrating
the latter over $k_T$,
\begin{eqnarray}
\phi_\pm(x)=\int d^2k_T\phi_\pm(x,k_T)\;.
\end{eqnarray}
The dependence of $\phi_{\pm}(x)$ and of $\phi_{\pm}(x,k_T)$ on
the renormalization scale $\mu$ has been suppressed.

Define the moments of the $B$ meson distribution amplitude
$\phi_+(x)$,
\begin{eqnarray}
\Lambda_0\equiv\int
dx\frac{\phi_+(x)}{x}\;,\;\;\;\;\Lambda_1\equiv\int
dx\phi_+(x)\;.\label{fac1}
\end{eqnarray}
The asymptotic behavior of $\phi_+(x)$ has been extracted from a
renormalization-group equation, which exhibits a decrease slower
than $1/x$ \cite{Neu03,BIK}. That is, the normalization
$\Lambda_1$ of the $B$ meson distribution amplitude is divergent
after taking into account the evolution effect. It has been argued
that a non-normalizable $B$ meson distribution amplitude does not
cause a trouble in practice \cite{OL}, since only the inverse
moment $\Lambda_0$ matters at LP \cite{JHEP042003029,BBNS}, which
is convergent. Note that a hard kernel would not be as simple as
$1/x$ at higher orders in $\alpha_s$, and information of more
moments is also necessary. In the following discussion we shall
neglect the evolution effect, and assume that $\phi_+(x)$ is
normalized to unity, i.e., $\Lambda_1=1$. Since the $B$ meson
distribution amplitudes absorb soft dynamics, the light quark
momentum $k$ is of $O(\bar\Lambda)$. We then have the relative
importance $\Lambda_1/\Lambda_0\sim \bar\Lambda/m_b$ for $x\sim
O(\bar\Lambda/m_b)$.

The factorization of soft dynamics from the transition matrix
element associated with the second term on the right-hand side of
Eq.~(\ref{cur}) gives the nonlocal matrix element,
\begin{eqnarray}
\langle 0|{\bar u}_\rho(z) i\slashs D h_{\delta}(0)|\bar
B(P_1)\rangle\;.\label{m2}
\end{eqnarray}
The factorization of the transition matrix elements with the
insertion of the $O(1/m_b)$ interactions in Eq.~(\ref{o12}) into
Eq.~(\ref{lm}) leads to,
\begin{eqnarray}
\langle 0|i\int d^4yT[{\bar u}_\rho(z)
h_{\delta}(0)O_{1,2}(y)]|\bar B(P_1)\rangle\;.\label{m4}
\end{eqnarray}
The contributions from Eqs.~(\ref{m2}) and (\ref{m4}) can be
absorbed into the nonlocal matrix element,
\begin{eqnarray}
\langle 0|{\bar u}_{\rho}(z)b_{\delta}(0)|\bar
B(P_1)\rangle\;,\label{bwr}
\end{eqnarray}
where the rescaled $b$ quark field $h$ has been replaced by the
full field $b$. It is easy to check that the heavy-quark expansion
of Eq.~(\ref{bwr}) generates Eqs.~(\ref{m2}) and (\ref{m4}). This
absorption makes sense, because Eqs.~(\ref{m2}) and (\ref{m4}),
concerning only the initial $b$ quark, are universal for all
exclusive $B$ meson decays. The decomposition in Eq.~(\ref{bwp2})
still holds, but the $B$ meson distribution amplitudes $\Phi_\pm$,
redefined by Eq.~(\ref{bwr}) in terms of the full field $b$,
exhibit a renormalization-group evolution different from that in
Eq.~(\ref{bwp2}) \cite{LL04}.

\subsection{Three-parton Distribution Amplitudes}

\begin{figure}[t!]
\begin{center}
\epsfig{file=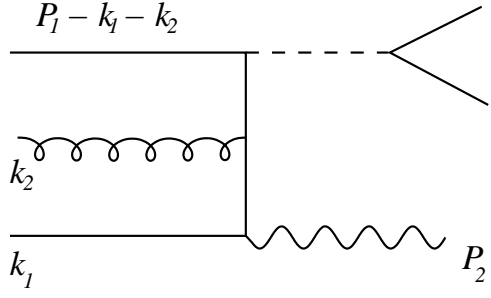,height=1.5in}
\end{center}
\caption{Three-parton contribution to the $B\to\gamma l\nu$ decay.
\label{fig4}}
\end{figure}

We explain that the nonlocal matrix element in Eq.~(\ref{m3}) is
negligible in the current accuracy: the three-parton distribution
amplitudes, whose contributions to the form factors are supposed
to be of $O(1/m_b)$, turn out to appear at $1/m_b^2$. The
three-parton distribution amplitudes $\tilde{\Phi}_V$,
$\tilde{\Phi}_A$, $\tilde{X}_A$, and $\tilde{Y}_A$ in coordinate
space are defined via the decomposition,
%\begin{eqnarray}
% \lefteqn{\langle 0 | \bar{u}_\rho(z) g G_{\alpha\beta} (uz)z^\beta
% h_{\delta}(0) | \bar{B}(P_1) \rangle}\nonumber \\
%  &=& f_B \left\{
%        (\slashs P_1 + m_B)\gamma_5\biggl[ ( v_{\alpha}\slashs z
%         - t \gamma_{\alpha} )  \left(\tilde{\Phi}_V (t,u) -
%          \tilde{\Phi}_A (t,u)\right)
%      \right. \nonumber\\
%  & & - i \sigma_{\alpha\beta}z^\beta\tilde{\Phi}_V (t,u)
%  \left. - z_{\alpha}\tilde{X}_A (t,u) -
%\frac{z_\alpha}{t}\slashs z \tilde{Y}_A(t,u) \biggr]
%\right\}_{\delta\rho}\;,
% \lefteqn{\langle 0 | \bar{u}_\rho(z) g G_{\alpha\beta} (uz)
% h_{\delta}(0) | \bar{B}(P_1) \rangle}\nonumber \\
%  &=& f_B \left\{
%        (\slashs P_1 + m_B)\gamma_5\biggl[ (
%        v_{\alpha}\gamma_\beta
%         - v_\beta \gamma_{\alpha} )  \left(\tilde{\Phi}_V (t,u) -
%          \tilde{\Phi}_A (t,u)\right)
%      \right. \nonumber\\
%  & & - i \sigma_{\alpha\beta}\tilde{\Phi}_V (t,u)
%+\left. \frac{1}{t}(v_\alpha z_\beta-v_\beta z_{\alpha})
%\tilde{X}_A (t,u) +
%\frac{1}{t}(z_\alpha\gamma_\beta-z_\beta\gamma_\alpha)
%\,\tilde{Y}_A \,(t,u)\biggr] \right\}_{\delta\rho} \;,
%\label{3elements}
%\end{eqnarray}
%with the variable $t=v\cdot z$. For the application to the
%$B\to\gamma l\nu$ decay, we consider
\begin{eqnarray}
 \lefteqn{\langle 0 | \bar{u}_\rho(z) g G_{\alpha\beta} (uz)
 n_-^\beta h_{\delta}(0) | \bar{B}(P_1) \rangle}\nonumber \\
 &=&f_B \left\{
        (\slashs P_1 + m_B)\gamma_5\biggl[ (
        v_{\alpha}\slashs n_-
         - v\cdot n_- \gamma_{\alpha} )  \left(\tilde{\Phi}_V (t,u) -
          \tilde{\Phi}_A (t,u)\right)
      \right. \nonumber\\
  & & - i \sigma_{\alpha\beta}n_-^\beta \tilde{\Phi}_V (t,u)
\left.- n_{-\alpha} \tilde{X}_A (t,u) + \frac{n_{-\alpha}}{v\cdot
n_-}\slashs n_- \,\tilde{Y}_A \,(t,u) \biggr] \,
\right\}_{\delta\rho}\;, \label{3e2}
\end{eqnarray}
with the variable $t=v\cdot z$. The corresponding hard kernels
arise from the contraction of all the structures
$\Gamma=v_{\alpha}\slashs n_-$, $v\cdot n_- \gamma_{\alpha}$,...,
in Eq.~(\ref{3e2}) with Fig.~\ref{fig4}, written as
\begin{eqnarray}
{\cal M}_a^{(3)}\propto \frac{tr\{\slashs \epsilon_T^* [u\slashs
n_+\gamma^\alpha(\slashs P_2-\slashs k_1-u\slashs k_2) -\bar
u(\slashs P_2-\slashs k_1-u\slashs k_2)\gamma^\alpha\slashs
n_+]\gamma_\mu(1-\gamma_5)(\slashs
P_1+m_B)\gamma_5\Gamma\}}{[(P_2-k_1-uk_2)^2]^2}\;, \label{h2}
\end{eqnarray}
where $k_1$ ($k_2$) is the momentum carried by the light quark
(gluon). The derivation of the above expression is referred to the
Appendix.

%For the application to the $B\to\gamma l\nu$ decay, we choose
%$\Gamma=\gamma_5\gamma_\delta$ and consider
%\begin{eqnarray}
% \lefteqn{\langle 0 | \bar{q} (z) \, g G_{\alpha\beta}
% (uz)n_-^\alpha
%      \, \gamma_5\gamma_\delta \, b (0) | \bar{B}(P_1) \rangle}\nonumber \\
%  &=& f_B m_B\, n_-^\alpha\left[ \, ( v_{\alpha}g_{\delta\beta}
%         - v_\beta g_{\delta\alpha} )\  \tilde{\Psi}_A (t,u)
%      \right. \label{3s}\\
%  & & +\left. \frac{v_\delta}{v\cdot n_-}(v_\alpha n_{-\beta}
%  -v_\beta n_{-\alpha}) \, \tilde{X}_A (t,u)\,
%+ \frac{1}{v\cdot n_-}(n_{-\alpha}g_{\delta\beta}
%-n_{-\beta}g_{\delta\alpha}) \,\tilde{Y}_A \,(t\,,\,u)  \,
%\right]\;, \nonumber\\
%&=&f_B m_B\, \left[ \, ( v\cdot n_-g_{\beta\delta}
%         - v_\beta n_{-\delta} )\  \tilde{\Psi}_A (t,u)
%+ n_{-\beta}v_\delta \, \tilde{X}_A (t,u)\,
%-\frac{n_{-\beta}n_{-\delta}}{v\cdot n_-} \,\tilde{Y}_A \,(t,u) \,
%\right]\;. \label{3e1}
%\end{eqnarray}

For $\Gamma=v\cdot n_- \gamma_{\alpha}$, Eq.~(\ref{h2}) vanishes
because of $\epsilon_T^*\cdot n_+=\epsilon_T^*\cdot (P_2-
k_1-uk_2)=0$. Express
$\sigma_{\alpha\beta}n_-^\beta=i(n_{-\alpha}-\slashs
n_-\gamma_\alpha)$, in which the first term has the same structure
as of $\tilde{X}_A$. The second term $\slashs n_-\gamma_\alpha$
renders Eq.~(\ref{h2}) vanish for the same reason. For the other
structures $v_\alpha \slashs n_{-}$, $n_{-\alpha}$, and
$n_{-\alpha}\slashs n_{-}$, we always have
$\gamma^\alpha=\gamma^+$. Once $\gamma^\alpha=\gamma^+$,
Eq.~(\ref{h2}) is proportional to
\begin{eqnarray}
{\cal M}_a^{(3)}\sim\frac{P_1\cdot (k_1+uk_2)}{[(P_2\cdot
(k_1+uk_2)]^2}\;. \label{h23}
\end{eqnarray}
Note that $k_1^+$ and $k_2^+$ are of $O(\bar\Lambda)$, and that
the moments of the three-parton $B$ meson distribution amplitudes
are at most of $O(\bar\Lambda^2)$ \cite{GN}. Therefore, when
convoluting Eq.~(\ref{h23}) with the three-parton distribution
amplitudes, the resultant contribution to the form factors
$F_{V,A}$ is of $O(\bar\Lambda^2/m_b^2)$ compared to the LP one
from Fig.~\ref{fig1}(a). The same higher Fock state has been shown
to give a power-suppressed correction to the $B\to\gamma l\nu$
decay in the framework of soft-collinear effective theory (SCET)
\cite{BHLN}. With a similar reasoning, the three-parton $B$ meson
wave functions also contribute at $O(1/m_b^2)$ in $k_T$
factorization theorems. We emphasize that the three-parton $B$
meson wave functions are relevant in the NLP analysis of the
$B\to\pi$ transition form factors. This is the reason the
$B\to\gamma l\nu$ decay is a cleaner mode than the $B\to\pi l\nu$
decay for determining the LP $B$ meson wave function.

\subsection{NLP Hard Kernels}

Contracting Fig.~\ref{fig1} with the two structures in
Eq.~(\ref{bwp2}), we get the quark-level amplitudes,
\begin{eqnarray}
{\cal M}^{+}_{a}&=&\frac{i}{4\sqrt{2}} \frac{tr[\slashs
\epsilon^*(\slashs P_2-\slashs k)\gamma_\mu(1-\gamma_5) (\slashs
P_1+m_B)\gamma_5\slashs n_+]}{(P_2-k)^2}\;,\nonumber\\
{\cal M}^{+}_{b}&=&\frac{i}{4\sqrt{2}}
\frac{tr[\gamma_\mu(1-\gamma_5)(\slashs q-\slashs k +m_b) \slashs
\epsilon^*(\slashs P_1+m_B)\gamma_5\slashs n_+]}{(q-k)^2-m_b^2}\;,
%H^{-}_{a}(x,q^2)&=&\frac{i}{4\sqrt{2}} \frac{tr[\slashs
%\epsilon^*(\slashs P_2-\slashs k)\gamma_\mu(1-\gamma_5) (\slashs
%P_1+m_B)\gamma_5 \slashs n_-]}{(P_2-k)^2}\;,\\
%H^{-}_{b}(x,q^2)&=&\frac{i}{4\sqrt{2}}
%\frac{tr[\gamma_\mu(1-\gamma_5)(\slashs q-\slashs k +m_b) \slashs
%\epsilon^*(\slashs P_1+m_B)\gamma_5\slashs
%n_-]}{(q-k)^2-m_b^2}\;.
\label{hva}
\end{eqnarray}
and ${\cal M}^{-}_{a,b}$ with the null vector $n_+$ in ${\cal
M}^{+}_{a,b}$ being replaced by $n_-$. As stated above,
Fig.~\ref{fig1}(a) is LP, because of $(P_2-k)^2=-2P_2\cdot k\sim
O(m_b\bar\Lambda)$, and Fig.~\ref{fig1}(b) is NLP, because of
$(q-k)^2-m_b^2=-2P_1\cdot P_2\sim O(m_b^2)$. The contribution from
Fig.~\ref{fig1}(b) has not yet been considered in the literature.
We shall neglect the mass difference between the $B$ meson and the
$b$ quark in ${\cal M}_b^{+,-}$ in our analysis accurate up to
NLP.

%$H_V$, $H_A$, ${\bar H}_V$, and ${\bar H}_A$  are written as
%\begin{eqnarray}
%H_V&=&\sqrt{\frac{N_c}{2}} \frac{4}{\eta
%m_B}\left(\frac{1}{x}+\frac{1}{\eta}+\frac{1}{1-x}-\frac{x}{\eta(1-x)}\right)\;,\nonumber\\
%H_A&=&\sqrt{\frac{N_c}{2}} \frac{4}{\eta
%m_B}\left(\frac{1}{x}-\frac{1}{\eta}-\frac{1}{1-x}-\frac{x}{\eta(1-x)}\right)\;,\nonumber\\
%{\bar H}_V&=&-\sqrt{\frac{N_c}{2}} \frac{4}{\eta^2
%m_B}\left(1-\frac{x}{1-x}\right)\;,\nonumber\\
%{\bar H}_A&=&\sqrt{\frac{N_c}{2}} \frac{4}{\eta^2
%m_B}\frac{1}{1-x}\;.
%\end{eqnarray}
%Since $x\sim O(\bar\Lambda/m_B)$, we expand the factor
%$1/(1-x)\sim 1$.

The collinear factorization formulas for $F_{V,A}$ are written as
\begin{eqnarray}
F_{V(A)}(q^2)&=&f_B\int dx \left[\phi_+(x)H_{V(A)}^{+}(x,\eta)+
\phi_-(x)H_{V(A)}^{-}(x,\eta)\right]\;,\label{fva}
\end{eqnarray}
where the hard kernels $H$ are extracted according to
Eq.~(\ref{ME1b}) by keeping only the longitudinal component $k^+$
in Eq.~(\ref{hva}). In terms of the LP and NLP moments in
Eq.~(\ref{fac1}), Eq.~(\ref{fva}) becomes
\begin{eqnarray}
F_{V,A}(q^2)=\frac{f_B}{\eta
m_B}\left[\Lambda_0\pm\left(1+\frac{1}{\eta}\right)\Lambda_1\right]\;,
%\nonumber\\
%F_{A}(q^2)&=&\frac{f_B}{\eta
%m_B}\left[\Lambda_0-\left(1+\frac{1}{\eta}\right)\Lambda_1\right]\;.
\label{dcf}
\end{eqnarray}
in which the coefficient 1 of $\Lambda_1$ comes from
Fig.~\ref{fig1}(a) and $1/\eta$ from Fig.~\ref{fig1}(b).  It has
been mentioned that the equality of $F_V$ and $F_A$ at LP is
attributed to the spin symmetry in the large-recoil region
\cite{PRD61-114510}. The coefficient $1/\eta$ implies the increase
of the subleading-power correction with the decrease of the photon
energy. This is why a perturbation theory is reliable only in the
large $\eta$ region. The distribution amplitude $\phi_-(x)$,
contributing only through the normalization of the combination,
\begin{eqnarray}
\int dx[\phi_+(x)-\phi_-(x)]=0\;,
\end{eqnarray}
disappears from Eq.~(\ref{dcf}). As shown in Eq.~(\ref{dcf}), the
normalization $\Lambda_1$ does appear at NLP, which is divergent
under the evolution. This is another example that the QCD-improved
factorization (QCDF) approach based on collinear factorization
theorem breaks down at NLP \cite{BBNS,Monr}.

The decay spectrum in Eq.~(\ref{d11}) becomes
\begin{eqnarray}
\frac{d\Gamma }{d\eta}&=&\frac{\alpha G_F^2\left| {V}_{ub}\right|
^2}{48\pi^2}f_B^2m_B^3
(1-\eta)\eta\left[\Lambda_0^2+\left(1+\frac{1}{\eta}\right)^2\Lambda_1^2\right]\;.
\label{d1}
\end{eqnarray}
The above expression indicates that the NLP terms for the spectrum
have cancelled, and only the $O(1/m_b^2)$ term $\Lambda_1^2$ is
left. In this case we can estimate the $O(1/m_b^2)$ effect using
the models for the $B$ meson distribution amplitudes available in
the literature \cite{KKQT,HWZ},
\begin{eqnarray}
\phi_{\pm}(x)=\frac{\lambda\pm
(x-\lambda)}{2\lambda^2}\theta(x)\theta(2\lambda-x)\;, \label{KK}
\end{eqnarray}
with the shape parameter $\lambda\equiv \bar\Lambda/m_b$. The
value of $\bar\Lambda$ has been found to range between 0.5 and 0.7
GeV \cite{JHEP042003029,WY,HW04}, which corresponds to
$\lambda=0.1\sim 0.15$ approximately. Certainly, there are other
models of the $B$ meson distribution amplitudes (see \cite{LRev}).

\begin{figure}[t!]
\begin{center}
\epsfig{file=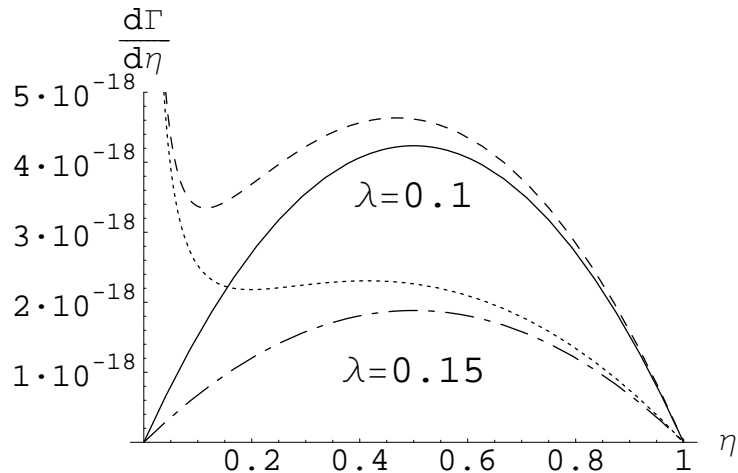,height=2.5in}
\end{center}
\caption{Spectra Spectra in units of GeV$^{-1}$ from collinear
factorization with the solid (dashed-dotted)line corresponding to
the LP contribution for $\lambda=0.1$ ($\lambda=0.15$), and the
dashed (dotted)line to the inclusion of the NLP contribution for
$\lambda=0.1$ ($\lambda=0.15$).}\label{qcdf}
\end{figure}

Employing the inputs $\alpha=1/137$, $G_F=1.16639\times 10^{-5}$
GeV$^{-2}$, $|V_{ub}|=3.9\times 10^{-3}$, $f_B=190$ MeV, and $m_B
= 5.28$ GeV, we derive the photon energy spectra of the
$B\to\gamma l\nu$ decay for $\lambda=0.1$ and for $\lambda=0.15$
in Fig.~\ref{qcdf}. The specific models in Eq.~(\ref{KK}) lead to
the relation $\Lambda_1/\Lambda_0=\lambda$. Therefore, the
subleading-power term is indeed negligible at large $\eta$, whose
contribution is around 5\%. However, this term diverges quickly at
small $\eta$, breaking the perturbative expansion in $1/m_b$. The
form factors $F_{V,A}$ in Eq.~(\ref{dcf}) contain a dominant
monopole component proportional to $\Lambda_0/\eta$, and a small
dipole component proportional to $\Lambda_1/\eta^2$, which is
important only at small $\eta$. This is the reason one always
obtains a symmetric spectrum in $\eta$ at LP from a perturbation
theory \cite{PRD61-114510} as shown in Fig.~\ref{qcdf}. To
generate an asymmetric spectrum, the dipole component must be
enhanced as postulated in \cite{PLB361-137,KAG}. Therefore, an
asymmetric spectrum signals an important NLP contribution, ie., a
breakdown of factorization theorem.
%The issue of the shape of the
%$B\to\gamma l\nu$ spectrum was also discussed in
%\cite{PRD64-097503}.

It has been explained that the undesirable feature of the $B$
meson distribution amplitude under evolution is a consequence of
collinear factorization, which can be removed in $k_T$
factorization \cite{LL04}. The evolution effect on the
$k_T$-dependent $B$ meson wave function was also studied in
\cite{JMW04}. Moreover, applying $k_T$ factorization theorem to
the $B\to\gamma l\nu$ decay, which has been proved in \cite{NL2},
we can extend the spectrum to lower $\eta$ as demonstrated below.
%In this formalism, the hard kernels are modified into
%\begin{eqnarray}
%H_V&=&\sqrt{\frac{N_c}{2}}4m_B\left(\frac{1}{\eta xm_B^2+k_T^2}
%+\frac{x}{\eta(\eta xm_B^2+k_T^2)}+\frac{1}{\eta (1-x)m_B^2+k_T^2}
%-\frac{x}{\eta[\eta (1-x)m_B^2+k_T^2]}\right)\;,\nonumber\\
%H_A&=&\sqrt{\frac{N_c}{2}}4m_B\left(\frac{1}{\eta xm_B^2+k_T^2}
%-\frac{x}{\eta(\eta xm_B^2+k_T^2)}-\frac{1}{\eta (1-x)m_B^2+k_T^2}
%-\frac{x}{\eta[\eta (1-x)m_B^2+k_T^2]}\right)\;,\nonumber\\
%{\bar H}_V&=&-\sqrt{\frac{N_c}{2}}
%4m_B\frac{x}{\eta}\left(\frac{1}{\eta
%xm_B^2+k_T^2}-\frac{1}{\eta (1-x)m_B^2+k_T^2}\right)\;,\nonumber\\
%{\bar H}_A&=&\sqrt{\frac{N_c}{2}}
%4m_B\frac{x}{\eta}\left(\frac{1}{\eta xm_B^2+k_T^2}+\frac{1}{\eta
%(1-x)m_B^2+k_T^2}\right)\;.
%\end{eqnarray}
Keeping both the longitudinal momentum $k^+$ and the transverse
momentum $k_T$ in Eq.~(\ref{hva}), the hard kernels in $k_T$
factorization theorem are derived. Defining the LP and NLP
functions,
\begin{eqnarray}
\Lambda_0(\eta)&\equiv&m_B^2\int dx\int
d^2k_T\frac{\phi_+(x,k_T)}{\eta
xm_B^2+k_T^2}\;,\nonumber\\
\Lambda_1(\eta)&\equiv&m_B^2\int dx\int
d^2k_T\left[\frac{\phi_+(x,k_T)}{\eta m_B^2+k_T^2}
+\frac{x\phi_-(x,k_T)}{\eta(\eta xm_B^2+k_T^2)}
\right]\;,\label{fac2}
%\Lambda_0(\eta)&\equiv&m_B^2\int dx\int_0^{\infty}
%bdb\phi_+(x,b)K_0\left(\sqrt{\eta
%x}m_Bb\right)\;,\nonumber\\
%\Lambda_1(\eta)&\equiv&m_B^2\int dx\int_0^{\infty}
%bdb\left[\phi_+(x,b)K_0\left(\sqrt{\eta}m_Bb\right)
%+\phi_-(x,b)\frac{x}{\eta}K_0\left(\sqrt{\eta x}m_Bb\right)
%\right]\;,\label{fac2}
\end{eqnarray}
respectively, we obtain the form factors,
\begin{eqnarray}
F_{V,A}(q^2)
%&=&\sqrt{\frac{N_c}{2}} 4m_B\int dx\int_0^{\infty}
%bdb\left\{\phi_B(x,b)
%\left[\left(1+\frac{x}{\eta}\right)K_0\left(\sqrt{\eta
%x}m_Bb\right) +K_0\left(\sqrt{\eta }m_Bb\right)\right]\right.
%\nonumber\\
%& &\hspace{4.0cm}
%\left.-\bar\phi_B(x,b)\frac{x}{\eta}K_0\left(\sqrt{\eta
%x}m_Bb\right)\right\}\;,\nonumber\\
=\frac{f_B}{m_B}\left[\Lambda_0(\eta)\pm\Lambda_1(\eta)\right]\;.
%F_{A}(q^2)
%&=&\sqrt{\frac{N_c}{2}} 4m_B\int dx\int_0^{\infty}
%bdb\left\{\phi_B(x,b)
%\left[\left(1-\frac{x}{\eta}\right)K_0\left(\sqrt{\eta
%x}m_Bb\right) -K_0\left(\sqrt{\eta }m_Bb\right)\right]\right.
%\nonumber\\
%& &\hspace{4.0cm}
%\left.+\bar\phi_B(x,b)\frac{x}{\eta}K_0\left(\sqrt{\eta
%x}m_Bb\right)\right\}\;,\nonumber\\
%&=&
%\frac{f_B}{m_B}\left[\Lambda_0(\eta)-\Lambda_1(\eta)\right]\;,
\label{fkt}
\end{eqnarray}
Because of $k_T\sim O(\bar\Lambda)$ in the $B$ meson,
$\Lambda_1(\eta)$ is of $O(\bar\Lambda/m_b)$ relative to
$\Lambda_0(\eta)$ in the large $\eta$ region. Again, only a single
$B$ meson wave function is relevant in the LP analysis of the
$B\to \gamma l\nu$ decay, consistent with the observation in
\cite{TLS}. Compared to Eq.~(\ref{dcf}), both $\phi_\pm$ appear in
the $k_T$ factorization theorem at NLP.

\begin{figure}[t!]
\begin{center}
\epsfig{file=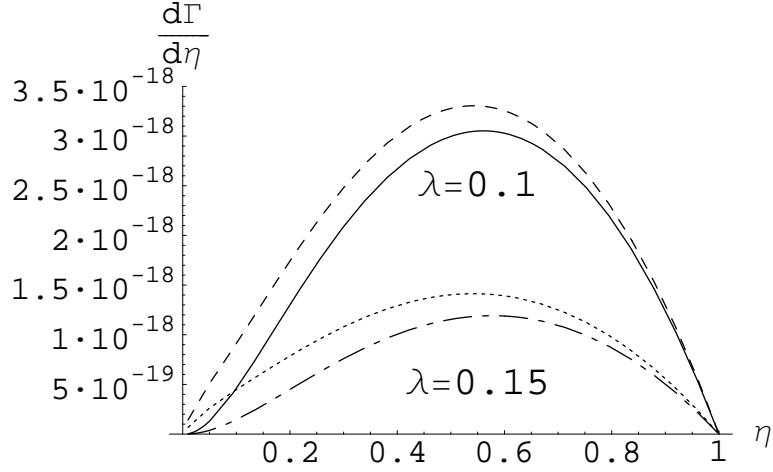,height=2.5in}
\end{center}
\caption{Spectra Spectra in units of GeV$^{-1}$ from $k_T$
factorization with the solid (dashed-dotted)line corresponding to
the LP contribution for $\lambda=0.1$ ($\lambda=0.15$), and the
dashed (dotted)line to the inclusion of the NLP contribution for
$\lambda=0.1$ ($\lambda=0.15$).}\label{pqcd}
\end{figure}

The decay spectrum is then given, according to Eq.~(\ref{d11}), by
\begin{eqnarray}
\frac{d\Gamma }{d\eta}=\frac{\alpha G_F^2\left| {V}_{ub}\right| ^2
}{48\pi ^2}f_B^2m_B^3
(1-\eta)\eta^3\left[\Lambda_0^2(\eta)+\Lambda_1^2(\eta)\right]\;.\label{dkt}
\end{eqnarray}
Similarly, the NLP terms have cancelled, and only the
$O(\bar\Lambda^2/m_b^2)$ term $\Lambda_1^2(\eta)$ is left. We
adopt the models for the $B$ meson wave functions in \cite{KKQT},
whose $k_T$ dependence is coupled to the $x$ dependence through a
$\delta$-function,
\begin{eqnarray}
\phi_{\pm}(x,k_T)=\phi_{\pm}(x)
\frac{1}{\pi}\delta\left(k_T^2-x(2\lambda-x)m_B^2\right)\;.
%J_0\left(m_Bb\sqrt{x(2\lambda-x)}\right)\;.
\label{KKb}
\end{eqnarray}
%Equation (\ref{fac2}) then reduces to
%\begin{eqnarray}
%\Lambda_0(\eta)&\equiv&\int \frac{dx}{x}\frac{\phi_+(x)}{\eta
%+2\lambda-x}\;,\nonumber\\
%\Lambda_1(\eta)&\equiv&\int dx\left[\frac{\phi_+(x)}{\eta
%+x(2\lambda-x)} +\frac{\phi_-(x)}{\eta(\eta +2\lambda-x)}
%\right]\;.\label{fac3}
%\end{eqnarray}
Using the same input parameters, we obtain the photon energy
spectra from $k_T$ factorization theorem in Fig.~\ref{pqcd} for
$\lambda=0.1$ and for $\lambda=0.15$. These spectra are symmetric
in $\eta$, and modified only slightly by the higher-power
correction. Hence, the higher-power correction is under control in
$k_T$ factorization theorem compared to that in collinear
factorization theorem: the power behavior $1/\eta$ of the spectrum
in the small $\eta$ region has been smeared into $\eta\ln^2\eta$.
It implies that the perturbative QCD (PQCD) approach based on
$k_T$ factorization theorem \cite{LY1,CL,KLS,LUY} has a better
convergence at subleading level.

\section{LONG- AND SHORT-DISTANCE CORRECTIONS}

In this section we discuss the long-distance and short-distance
corrections to the $B\to\gamma l\nu$ decay spectrum. For this
purpose, the form factors are written, in $k_T$ factorization
theorem, as
\begin{eqnarray}
F_{V,A}(q^2)&=&\frac{f_B}{m_B}\left[\Lambda_0(\eta)
+\Lambda_0^{(1)}(\eta)\right]+F_{V,A}^{\rm LD}(q^2)\;,
\label{ffmsing}
\end{eqnarray}
where $\Lambda_0^{(1)}$ and $F_{V,A}^{\rm LD}$ denote the
$O(\alpha_s)$ and long-distance correction to the leading result,
respectively. We shall estimate the latter by considering the
$B\to V\to \gamma$ transition. This correction is certainly
significant in the small $\eta$ (large $q^2$) region, where the
internal quark becomes soft, and easily form a resonance with the
spectator quark. Hence, it could break the QCD factorization of
the form factors $F_{V,A}$ at small $\eta$. At large $\eta$, the
long-distance contribution may be suppressed by the values of the
$B\to V$ transition form factors \cite{PLB358-329}.

%Another piece of
%long-distance contribution through $B\to \gamma B^*\to \gamma
%l\nu$ will not be considered here, since it is a correction to the
%power-suppressed amplitude from Fig.~1(b).

The long-distance amplitude is written as \cite{KMS}
\begin{eqnarray}
\frac{1}{e}\left\langle \gamma \left( P_2,\epsilon_T\right) \left|
\bar{u}\gamma_\mu (1-\gamma_5)b\right| \bar B\left( P_1\right)
\right\rangle&=&\sum_V\left\langle\gamma\left(
P_2,\epsilon_T\right) | J_{em}^\alpha|V\left(
P_2,\epsilon_T\right)\right
\rangle\frac{-i\epsilon_{T\alpha}^*}{P_2^2-m_V^2+im_V\Gamma_V}\nonumber\\
& &\times \left\langle V\left( P_2,\epsilon_T\right) \left|
\bar{u}\gamma_\mu (1-\gamma _5)b\right| \bar B\left( P_1\right)
\right\rangle\;,\label{long}
\end{eqnarray}
with the vector mesons $V=\rho,\omega,\cdots$, and their masses
$m_V$ and widths $\Gamma_V$. Take the $B$ meson transition into a
transversely polarized $\rho$ meson as an example, for which the
first matrix element on the right-hand side of Eq.~(\ref{long})
gives
\begin{eqnarray}
\langle\gamma(P_2,\epsilon_T)|J_{em}^\alpha|\rho(P_2,\epsilon_T)\rangle=-\frac{i}{2}m_\rho
f_\rho\epsilon_T^\alpha\;,\label{a1}
\end{eqnarray}
$f_\rho$ being the $\rho$ meson decay constant. The second matrix
element is decomposed into
\begin{eqnarray}
\langle\rho(P_2,\epsilon_T)|{\bar u}\gamma_\mu(1-\gamma_5) b|\bar
B(P_1)\rangle
=-\frac{2V(q^2)}{m_B+m_\rho}\epsilon_{\mu\nu\rho\sigma}
\epsilon^{*\nu}_{T} P_{1}^{\rho}
P_{2}^{\sigma}-i(m_B+m_\rho)A_1(q^2)\epsilon_{T\mu}^{*}\;,
\label{af}
\end{eqnarray}
with the $B\to\rho$ form factors $V(q^2)$ and $A_1(q^2)$.
Combining Eqs.~(\ref{a1}) and (\ref{af}), we extract from
Eq.~(\ref{long}),
\begin{eqnarray}
F_V^{\rm LD}(q^2)&=&\frac{
f_\rho}{m_\rho-i\Gamma_\rho}\frac{m_B}{m_B+m_\rho}V(q^2)\;,
\nonumber\\
F_A^{\rm LD}(q^2)&=&\frac{
f_\rho}{m_\rho-i\Gamma_\rho}\frac{(m_B+m_\rho)}{\eta
m_B}A_1(q^2)\;.\label{ld}
\end{eqnarray}
For the long-distance contribution through the $B\to\omega$
transition, we have the similar expressions to Eq.~(\ref{ld}), but
with the charge factor $1/2$ in Eq.~(\ref{a1}) being replaced by
$1/6$, and the appropriate replacement of the vector meson mass
and of the decay constant. The $B\to \psi$ transitions do not
contribute in this case.
%the charge factor $1/2$ is replaced by $2/3$, and the vector meson
%mass and decay constant are also replaced appropriately.

\begin{figure}[t!]
\begin{center}
\epsfig{file=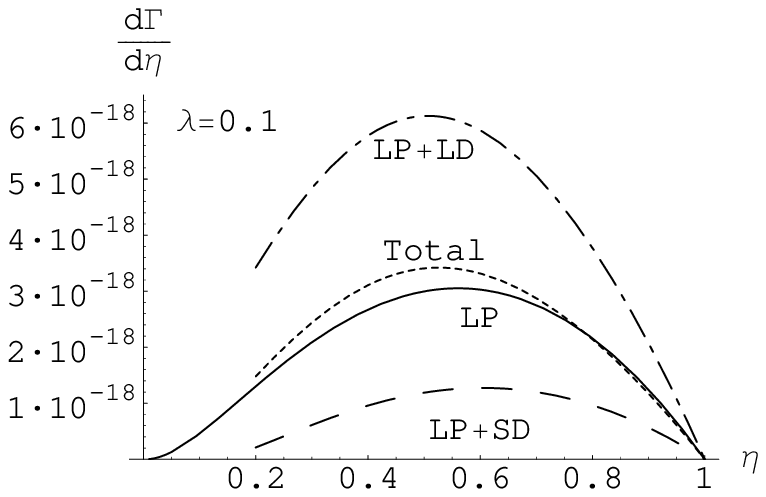,height=2.5in}
\epsfig{file=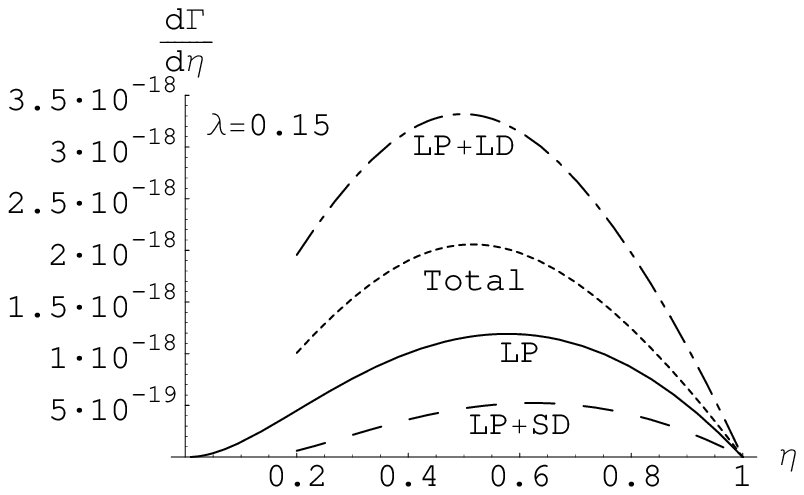,height=2.5in}
\end{center}
\caption{Spectra in units of GeV$^{-1}$ for $\lambda=0.1$ and
$\lambda=0.15$ with the solid lines corresponding to the LP
contribution only, the dashed-dotted lines to the inclusion of
long-distance contribution, the dashed lines to the inclusion of
the NLO correction, and the dotted lines to the inclusion of both
the long-distance and NLO contributions.}\label{sdld}
\end{figure}

For the $\rho$ and $\omega$ mesons, we employ the inputs
\cite{KMS}
\begin{eqnarray}
& &m_\rho=0.771\;\;{\rm
GeV}\;,\;\;\;\Gamma_\rho/m_\rho=0.21\;,\;\;\;f_\rho=0.217\;\;{\rm
GeV}\;,\nonumber\\
& & m_\omega=0.783\;\;{\rm
GeV}\;,\;\;\;\Gamma_\omega/m_\omega\approx
0\;,\;\;\;f_\omega=0.195\;\;{\rm GeV}\;.
\end{eqnarray}
For the $B\to \rho$, $\omega$ form factors, we adopt the models
derived from the light-front QCD \cite{CCH}, which have been
parameterized as
\begin{eqnarray}
F(q^2)=\frac{F(0)}{1-a(q^2/m_B^2)+b(q^2/m_B^2)^2}\;,\label{fp}
\end{eqnarray}
with the constants,
\begin{eqnarray}
V(q^2):& &F(0)=0.27\;,\;\;\;a=1.84\;,\;\;\;b=1.28\;,\nonumber\\
A_1(q^2):& &F(0)=0.22\;,\;\;\;a=0.95\;,\;\;\;b=0.21\;.
\end{eqnarray}
We restrict the above formalism in the region,
\begin{eqnarray}
\eta > 1-\frac{q^2_{\rm max}}{m_B^2}
%\equiv 1-\frac{(m_B-m_\omega)^2}{m_B^2}
=0.275\;,
\end{eqnarray}
$q^2_{\rm max}$ being the maximal value of $q^2$ in the
$B\to\omega$ transition, in which Eq.~(\ref{fp}) holds. The
long-distance contribution increases $F_{V,A}$ by about $30\sim
50$\% for $\lambda=0.1\sim$ 0.15 at large $\eta$, consistent with
the observation in \cite{PLB358-329,JHEP042003029,SO96}. Its effect to
the decay spectrum is quite important, especially for $\eta<0.8$,
as shown in Fig.~\ref{sdld}.

The $B\to\rho,\omega$ transition form factors at large recoil
could be regarded as an $O(\alpha_s)$ object \cite{TLS}. This
observation hints that we should attempt to take into account the
NLO short-distance correction to $F_{V,A}$. The NLO correction to
the $B\to\gamma l\nu$ decay has been computed by several groups
\cite{NPB649-349,BHLN,NPB650-356} in collinear factorization
theorem (SCET or QCDF). However, we need the result from $k_T$
factorization theorem (with the parton transverse momenta $k_T$
being included), which is quoted from \cite{PRD61-114510}:
\begin{eqnarray}
\Lambda_0^{(1)}(\eta)&=&-\frac{\alpha_s(2E_\gamma)}{4\pi} C_Fm_B^2
%\left[\frac{\alpha_s(m_b)}{\alpha_s(2E_\gamma)}\right]^{-2/\beta_0}
\int dx\int d^2k_T\frac{\phi_+(x,k_T)}{\eta
xm_B^2+k_T^2}\nonumber\\
& &\times\left[\ln^2\frac{\eta}{x} -\frac{5}{2}\ln\frac{\eta}{x}
+\frac{4\pi^2}{3}-\ln^2\left(1+\frac{k_T^2}{2k^{+2}}\right)+2\pi i
\ln\left(1+\frac{k_T^2}{2k^{+2}}\right) \right]\;.
\end{eqnarray}
%The factors $2\lambda-x$ in the above expression come from the
%integration of the $\delta$-function in Eq.~(\ref{KKb}) over
%$k_T^2$.
The weaker evolution of $f_B$ will be neglected for simplicity.
Due to the large negative double logarithm, the NLO correction to
the form factors $F_{V,A}$ is destructive, and about 30\% of the
leading result for both $\lambda=0.1$ and $\lambda=0.15$ at large
$\eta$. The resummation of this double logarithm to all orders has
been discussed in
\cite{PRD61-114510,NPB649-349,BHLN,NPB650-356,L5}.

We emphasize that the NLO hard kernel depends on a factorization
scheme, in which the $B$ meson wave function is defined
\cite{BHLN}. Therefore, it is not very legitimate to adopt an
expression straightforwardly from some other works in the
literature. The calculation of the NLO hard kernel for the
$B\to\gamma l\nu$ decay in the factorization scheme specified in
\cite{LL04} is in progress, which will be published elsewhere. The
NLO correction in SCET has been further factorized into a function
characterized by the scale $m_b$, and another by
$\sqrt{m_b\bar\Lambda}$. As stated in \cite{BHLN}, this further
factorization is not numerically essential for $m_b\approx 5$ GeV.
On the other hand, the model-dependent estimate of the
long-distance contribution also suffers large uncertainty. Hence,
we just intend to point out the potential strong cancellation
between the long-distance and short-distance corrections in this
mode. As shown in Fig.~\ref{sdld}, after combining both subleading
contributions, the net effect has been greatly reduced.
Especially, for the shape parameter $\lambda=0.1$, the
cancellation is almost exact for $\eta >0.8$. We conclude that the
leading result in the large $\eta$ region is stable under these
corrections.

Using the lifetime of a charged $B$ meson
$\tau_{B^\pm}=1.674\times 10^{-12}$ s and considering only the
leading contribution, we obtain the branching ratios for
$\lambda=0.15 \sim 0.1$,
\begin{eqnarray}
B(B\to\gamma l\nu)=(1.8\sim 4.8)\times 10^{-6}\;,\label{br}
\end{eqnarray}
from Eq.~(\ref{dkt}) in $k_T$ factorization theorem (PQCD), with
only the $O(\bar\Lambda^2/m_b^2)$ and $O(\alpha_s^2)$ uncertainty.
%Note that the spectrum in collinear factorization theorem (QCDF)
%leads to a logarithmically divergent total decay rate.
The values in Eq.~(\ref{br}) are more or less consistent with
other estimates in the literature: a model-dependent evaluation of
the structure-dependent photon emission contribution gave the
branching ratio $10^{-7}\sim 10^{-6}$ \cite{PRD51-111}. Using the
$B$ meson bound-state wave function from a Salpeter equation,
$0.9\times 10^{-6}$ has been obtained
\cite{PLB372-331,PRD64-097503}. Both a simple non-relativistic
model and light-front QCD led to $3.5\times 10^{-6}$
\cite{MPLA11-1061,PRD57-5697}.
%However, what was calculated in
%\cite{PRD57-5697} was considered as the soft (long-distance)
%contribution in \cite{JHEP042003029}, and the hard contribution
%analyzed here should be added.
Light-cone sum rules and the pole-model calculation gave $2\times
10^{-6}$ \cite{PLB361-137} and $2.26\times 10^{-6}$ \cite{KAG},
respectively. At last, the experimental upper bound at 90\%
confidence level was \cite{CLEO}
\begin{eqnarray}
B(B\to\gamma l\nu)<2.0\times 10^{-6}\;.
\end{eqnarray}

\section{CONCLUSION}

In this paper we have studied the $B\to\gamma l\nu$ decay in the
PQCD approach based on $k_T$ factorization theorem. This formalism
is well-defined at subleading level, since the two-parton LP $B$
meson wave functions remain normalizable even after including the
evolution effect. Note that the QCDF approach based on collinear
factorization theorem fails at NLP. We have shown that the
$O(1/m_b)$ corrections from the heavy-quark expansion can be
absorbed into the LP $B$ meson wave functions redefine by the
nonlocal matrix element in Eq.~(\ref{bwr}). The NLP contributions
from the hard kernels to the decay spectrum cancel. The
three-parton $B$ meson wave functions turn out to be suppressed by
$1/m_b^2$ in this special mode. The constructive long-distance
contribution almost cancels the destructive NLO radiative
correction for both the form factors $F_V$ and $F_A$. The $B$
meson wave function $\phi_+$ can then be extracted from the
observed $B\to\gamma l\nu$ decay spectrum using the leading
formalism, which suffers only the $O(1/m_b^2)$ and $O(\alpha_s^2)$
uncertainly. We conclude that the $B\to\gamma l\nu$ decay is the
cleanest mode for determining this important nonperturbative input
for the perturbation theories of exclusive $B$ meson decays. The
determination can be refined by including the evolution and
resummation effects into the factorization formulas
\cite{PRD61-114510,NPB649-349,BHLN,NPB650-356,L5}.

Measuring the $B\to \gamma l\nu$ spectrum in the lepton and photon
energies \cite{PRD61-114510},
\begin{eqnarray}
\frac{d^2\Gamma}{d \eta dy} &=& \frac{\alpha G_F^2 |V_{ub}|^2
m_B^3}{64\pi^2}(1-\eta)
\left\{ [F_V^2(q^2)+F_A^2(q^2)][2(1-y)(1-y-\eta)+\eta^2]\right.\nonumber\\
& &\left.\qquad\qquad -2F_V(q^2) F_A(q^2)
\eta(2-2y-\eta)\right\}\;,
\end{eqnarray}
with the lepton energy fraction $y=2E_l/m_B$, $1-\eta\le y\le 1$,
we can extract the information of the form factors $F_V$ and $F_A$
separately. It is then possible to fix the two two-parton $B$
meson wave functions $\phi_\pm$ simultaneously from
Eq.~(\ref{fkt}). At this NLP level, the three-parton wave
functions are still absent following the reasoning in Sec.~II.B.
The long-distance contribution and the NLO corrections also cancel
each other as indicated in Eq.~(\ref{ffmsing}). With the $B\to
\gamma l\nu$ branching ratio around $10^{-6}$, the above
experimental determination is possible.

\vskip 0.5cm We thank I. Bigi, Y. Kwon, Z. Ligeti, M. Neubert, T.
Onogi, and A.I. Sanda for helpful discussions. HNL acknowledges
the hospitality of Nagoya University during his visit, where this
work was initiated. This work was supported by the National
Science Council of R.O.C. under Grant No. NSC-93-2112-M-001-014.

\appendix

\section{THREE-PARTON CONTRIBUTION}

We start with Eq.~(1.3) in Ref.~\cite{BB89}:
\begin{eqnarray}
G^{(1)}(z)=\int d^4 w\frac{i(\slashs z-\slashs
w)}{2\pi^2(z-w)^4}ig\slashs A(w)\frac{i\slashs w}{2\pi^2w^4}\;,
\end{eqnarray}
which describes the interaction of a quark with a gluon. In
momentum space the above expression becomes
\begin{eqnarray}
G^{(1)}(z)=\int \frac{d^4 l}{(2\pi)^4}\int \frac{d^4
k_2}{(2\pi)^4}e^{i(k_2+l)\cdot z}\frac{i(\slashs k_2+\slashs
l)}{(k_2+l)^2}\gamma^\alpha\frac{i\slashs l}{l^2}ig{\tilde
A}_\alpha(k_2)\;,
\end{eqnarray}
where $l$ ($k_2$) is the momentum carried by the incoming quark
(gluon). The Feynman parametrization gives
\begin{eqnarray}
G^{(1)}(z)=-\int du\int \frac{d^4 l}{(2\pi)^4}e^{il\cdot z}\int
\frac{d^4 k_2}{(2\pi)^4}e^{iuk_2\cdot z}\frac{(\slashs l+u\slashs
k_2)\gamma^\alpha(\slashs l-\bar u\slashs k_2)}{(l^2)^2}ig{\tilde
A}_\alpha(k_2)\;,
\end{eqnarray}
where the variable change $l+\bar u k_2\to l$, $\bar u\equiv 1-u$,
has been applied.

In the case we are considering, the gluon momentum $k_2$ is of
$O(\bar\Lambda)$, since the $B$ meson is dominated by soft
dynamics. We expand the above expression up to $O(k_2)$:
\begin{eqnarray}
G^{(1)}(x)&=&-\int du\int \frac{d^4 l}{(2\pi)^4}e^{il\cdot z}\int
\frac{d^4 k_2}{(2\pi)^4}e^{iuk_2\cdot z}\Bigg\{\frac{\slashs
l\gamma^\alpha\slashs l}{(l^2)^2}+ \frac{u\slashs
k_2\gamma^\alpha\slashs l}{(l^2)^2}-\frac{\bar u\slashs
l\gamma^\alpha\slashs k_2}{(l^2)^2}\Bigg\} ig{\tilde
A}_\alpha(k_2)\;,\nonumber\\
&=&-\int du\int \frac{d^4 l}{(2\pi)^4}e^{il\cdot z}
\Bigg\{\frac{\slashs l\gamma^\alpha\slashs
l}{(l^2)^2}igA_\alpha(uz)+ \frac{u\slashs n_+\gamma^\alpha\slashs
l-\bar u\slashs l\gamma^\alpha\slashs
n_+}{(l^2)^2}ig\partial_\beta A_\alpha(uz)n_-^\beta\Bigg\}
\;.\label{g1}
\end{eqnarray}
The first term on the right-hand side of Eq.~(\ref{g1}),
contributing to a phase factor \cite{BB89}, will be dropped. For
convenience, we work in the light-cone gauge $A^+=0$, in which the
second and third terms are rewritten as
\begin{eqnarray}
G^{(1)}(z)&=&i\int du gG_{\alpha\beta}(uz)n_-^\beta \int \frac{d^4
l}{(2\pi)^4}e^{il\cdot z} \frac{u\slashs n_+\gamma^\alpha\slashs
l-\bar u\slashs l\gamma^\alpha\slashs n_+}{(l^2)^2} \;.\label{g2}
\end{eqnarray}
It is clear that the field strength
$gG_{\alpha\beta}(uz)n_-^\beta$ can be factored together with the
rescaled $b$ quark field $h$ and the light quark field $\bar u$
into the nonlocal matrix element in Eq.~(\ref{3e2}). The integrand
depending on $l$ is then identified as the hard kernel in momentum
space for the three-parton contribution. Employing Eq.~(\ref{g2})
for Fig.~\ref{fig4}, and substituting $P_2-k_1-uk_2$ for $l$, we
obtain Eq.~(\ref{h2}).

\end{document}